\documentclass[aps,prb,preprint,showpacs,showkeys]{revtex4-1}

\usepackage{graphicx}
\usepackage{epsfig}
\usepackage{color}
\usepackage{amsfonts,amssymb}
\usepackage{mathbbol}

\newcommand{\diag}{{\rm diag}\,}
\newcommand{\tr}{{\rm tr}\,}
\begin{document}

\title{Non--equilibrium dynamics of a system with Quantum Frustration}

\author{ Heiner Kohler$^{(1)}$, Andreas Hackl$^{(2)}$, Stefan Kehrein$^{(3)}$}
\date{\today}
\affiliation{
$^{(1)}$ Instituto de Ciencias Materiales de Madrid, CSIC, 
C/ Sor Juana In{\'e}s de la Cruz 3,
28049 Madrid, Spain\\
$^{(2)}$ SAP, 
SAP Allee 45, 68789 St.~Leon--Rot, Germany\\
$^{(3)}$ Departement of Physics, Georg--August--Universit{\"a}t G{\"o}ttingen\\
Friedrich--Hund Platz 1, 37077 G{\"o}ttingen}
\email{hkohler@icmm.csic.es}

\begin{abstract}
Using  flow equations, equilibrium and non--equilibrium dynamics of a two--level system  are investigated, which couples via non--commuting components to 
two independent oscillator baths.  In equilibrium the two--level energy splitting is protected when the TLS is coupled symmetrically to both bath. 
A critical asymmetry angle separates the localized from the delocalized phase. 
 On the other hand, real--time  decoherence of a non--equilibrium initial state is
 for a generic initial state faster for a coupling to two baths than for a single bath. 
 
\end{abstract}

\maketitle
\section{Introduction}
Under  the notations of {\em frustration of decoherence}  or {\em quantum frustration} effects are subsumed which are ascribed 
to the competition and mutual cancellation of two environments, which couple to non--commuting observables of a central system.
The notion was coined in \cite{cas03} and the effect has since then been studied in a variety of systems, like a two--level system (TLS) coupled to two oscillator bath \cite{cas03,nov05,guo12}  or to two spin--baths \cite{rao08}, a harmonic oscillator coupled to two oscillator bath \cite{koh05,koh06b, cuc10}  in spin--lattices \cite{cuc08} or Josephson networks \cite{giu08}. Most notably it was proposed as cooling mechanism \cite{ere08}.  The relation to Kondo physics was already pointed out in \cite{cas03}. 
Certain phenomena occuring in the two channel Kondo model or in the Bose--Fermi--Kondo model can actually be interpreted in terms of quantum frustration \cite{ zhu02,zar02}.

In the model originally studied in \cite{cas03,nov05}, a TLS  with energy gap $\Delta$ couples 
linearly with its two transversal components to two independent baths.  It will be called 2BTLS in the following. 
The strength of the ohmic coupling is measured by two quantities $\gamma_{3}^{(1)}$ and  $\gamma_{2}^
{(2)}$ (assuming a magnetic field in $x$ direction, bath 1 couples to the $z$--component and  bath 2 the $y$--component). 
One remarkable result of Ref.~\cite{cas03} were the renormalization group (RNG) equations
\begin{eqnarray}
\label{renequations}
\frac{d\gamma_3^{(1)}}{dl} &=& -2\gamma_3^{(1)}\gamma_2^{(2)}-\gamma_3^{(1)}h^2\ , 
\nonumber\\
 \frac{d\gamma_2^{(2)}}{dl} &=& -2\gamma_3^{(1)}\gamma_2^{(2)}-\gamma_2^{(2)}h^2 
 \nonumber\\
 \frac{d h}{dl} & = &\left(1-\gamma_3^{(1)}-\gamma_2^{(2)}\right)h \ ,
\end{eqnarray}
where $dl=-d\ln \omega_c$ is the differential of the flow parameter and $h=\Delta/\omega_c$, where $\omega_c$ is the cutoff frequency of the bath modes.  If either $\gamma_3^{(1)}$ or $\gamma_2^{(2)}$ is zero, the  RNG equations of the single bath spin--Boson 
model \cite{bra82,cha82} are recovered which predict a Kosterlitz Thouless phase transition for $\gamma=1$.   For  $\gamma_3^{(1)}$ $=$ $\gamma_2^{(2)}$  the renormalization flow is different:  $h$ scales always to infinity, i.~e.~a phase transition never occurs, not 
even for arbitrary strong coupling. This is by now one of the most striking signature of quantum frustration. 

However the question whether for large couplings the delocalized phase at symmetric coupling is stable against asymmetries remains unanswered by 
the above RNG equations. They do not yield any estimate for  the renormalized energy gap $\Delta_r$, respectively  Kondo temperature in the delocalized phase.

The body of publications, mentioned above focuses on thermal equilibrium. But the question whether or not quantum coherence of a non--equilibrium initial state is protected by quantum frustration is crucial for possible applications.  
Time evolution of a spin in non--equilibrium can be more complicated than exponential decay predicted by Bloch equations \cite{sli96}. In particular an initially decoupled central system might on a very short time scale, called quantum Zeno--time,  
incur initial slips. This happens for instance to the  dissipative harmonic oscillator \cite{haa85}.  In this case short times decoherence is indeed enhanced by a second bath and only later effects of quantum frustration occur  \cite{koh05,koh06b} . 
 
 We address the above questions for the 2BTLS using the method of Hamiltonian flow equations. Flow equations were introduced in the early nineties by  G{\l}azek and Wilson \cite{gla93} and about the same time by Wegner \cite{weg94}. The method rests upon a 
continuous diagonalization of the Hamiltonian, details can be found in \cite{keh06}.  It was applied to the single bath spin--Boson model in \cite{keh96,keh96a,keh97,sta02}. In particular it proved to yield good results for the renormalized energy gap $\Delta_r$.  
 
In this work  a generalization of equations (\ref{renequations}) is derived analytically, which embraces any kind of coupling to two baths. Numerically $\Delta_r$ is calculated as a function of an asymmetry angle, called $\theta$, which varies from zero (single bath) to $
\pi/4$ (completely symmetric). Whereas for weak coupling there is little dependence on the asymmetry angle, as the coupling becomes stronger the dependence on the asymmetry becomes more and more important. A symmetric coupling protects the gap and prevents 
the KT--phase transition. Identifying the critical angle allows us to plot a phase diagram in  the $\gamma_3^{(1)}$--$\gamma_2^{(2)}$ plane, where the localized and the delocalized phases are separated by a critical line.
 

Using techniques developed recently  \cite{hack08,hack09}  we address the question whether decoherence of  a non--equilibrium initial state is protected by a second bath. The answer to this can not be given without a careful distinction about what is meant by 
quantum decoherence. In a folkloristic definition decoherence is the decay of the off--diagonal elements in some pointer basis and relaxation the decay of the diagonal elements. For a two--level systems both processes are obviously not independent and it is therefore 
not easy to distinguish them.  

For symmetric coupling we find that the moduli of off--diagonal elements in the eigenbasis of the spin operator in $x$--direction incur initial slips and subsequent oscillations on a time scale of the cutoff--frequency $\omega_c$. These initial slips on the time scale of 
$\omega_c$  are absent for a single baths, however the subsequent decay is oscillatory also in this case.  The expectation value of the spin operator in $x$ direction behaves quite differently. Here the decay is initially faster for a single bath but slows down on the time scale of the Rabi frequency $\Delta^{-1}$. On the other hand for symmetric coupling the decay is initially slow but increases later to reach an equilibrium value, which is smaller than for a single bath.   

In the first two sections of the manuscript we set up the model derive the flow equations and calculate equilibrium quantities. In section \ref{therm} the non--equilibrium dynamics is considered.

\section{Flow equations for the 2BTLS}

The Hamiltonian of the 2BTLS is given by 
\begin{eqnarray}
\label{ham}
H^{(0)} &=& H_0+ H_I^{(0)}\nonumber\\
H_0&=& - \Delta S_1+ \sum_{n=1}^2 \sum_k \omega^{(n)}_k a_{n,k}^\dagger a_{n,k}\\ 
H_{\rm I}^{(0)}&=& S_3\otimes \sum_k \lambda^{(1)}_{3,k} \left(a_{1,k}+a_{1,k}^\dagger\right)+ i S_2\otimes \sum_k \lambda^
{(2)}_{2,k} \left(a_{2,k}-a_{2,k}^\dagger\right)\nonumber
\end{eqnarray}
where $S_i$ are spin $\frac{1}{2}$--matrices and $a_{n,k}$ are bosonic annihilation operators $[a_{n,k},a_{m,k^\prime}^\dagger] $ $=$ $\delta_{kk^\prime}\delta_{nm}$ and
$[a_{n,k},a_{m,k^\prime}] $ $=$ $[a_{n,k}^\dagger,a_{m,k^\prime}^\dagger] $ $=$ $0$.
We will also use $S_0=\frac{1}{ 2}{\mathbb1}_2$.  The sum runs over the $N$ bath modes, where 
$N$ is assumed a large number such that the spectral functions
 \begin{eqnarray}
 \label{spectralfunc}
J_{3}^{(1)}(\omega)& =& \sum_k (\lambda_{3,k}^{(1)})^2\delta(\omega - \omega^{(1)}_k)\ , 
\nonumber\\ 
J_{2}^{(2)}(\omega) & =& \sum_l (\lambda_{2,l}^{(2)})^2\delta(\omega - \omega^{(2)}_l)
\end{eqnarray}
of both baths are smooth functions. They obey an Ohmic power law for small frequencies 
 $J_{3}^{(1)}(\omega)$ $ = $ $2 \gamma_{3}^{(1)} \omega$ 
and $J_{2}^{(2)}(\omega) $ $=$ $ 2 \gamma_{2}^{(2)}\omega$ and are regularized by a cutoff $
\omega_c \gg\Delta$. For simplicity we assume here and in the following the cut--off and the number 
of bath modes to be the same for both baths.

The Hamiltonian is approximately diagonalized by a unitary transformation \cite{weg94,keh06} 
which 
depends continuously on a flow parameter $B$. Any one--parameter family of unitarily equivalent 
Hamiltonians obeys the equation
\begin{equation}
\label{flow1}
\frac{d}{d B} H^{(0)}(B) = [\eta^{(0)}(B),H^{(0)}(B)]  
\end{equation}
with a properly chosen anti--Hermitian operator $\eta^{(0)}$. If $\eta^{(0)}$ is chosen as the 
commutator $ \eta^{(0)}$ $= $ $ [H_0(B), H_I^{(0)}(B)]$ it can be readily shown, see e.~g.~\cite
{weg06} that if $H_0$  is non--degenerate in the limit $B\to \infty$, $\tr H(B) H_I^{(0)}\to 0$ and thus 
the Hamiltonian 
becomes diagonal. The 
commutator on the right hand side of  equation (\ref{flow1})  generates interaction terms not present 
in $H^{(0)}$. They are formally included  in a more general Hamiltonian $H= H^{(0)}+ H^{(1)}$ and 
in a new generator $\eta = [ H_0, H]$. The equations are closed by neglecting normal ordered 
products of more than two creation or annihilation operators. 

 In order to write  the interacting part of the 
form invariant  Hamiltonian $H$ in a compact form it is useful to arrange the creation and annihilation operators in a $4 N$ vector 
$\vec{A\,}^T $ $= $ $(\vec{a_1}^T,(\vec{a_1\,}^\dagger)^T, \vec{a_2\,}^T,(\vec{a_2\,}^\dagger)^T )$, where 
$ \vec{a\,}_{n}^T$ $=$ $(a_{n,1},\ldots,a_{n,N})$,  $( \vec{a\,}^\dagger_{n})^T$ $=$ $(a^\dagger_{n,1},\ldots,a^\dagger_{n,N})$, $n=1,2$. It turns out useful as well to introduce coupling constants
$\lambda_{\pm,k}^{(n)}$ $\equiv$ $\lambda_{3,k}^{(n)} \pm \lambda_{2,k}^{(n)}$ and arrange them in a $4 N$ vector 
$\vec{\Lambda\,}$ = $(\vec{\lambda\,}_+^{(1) T}, \vec{\lambda\,}_-^{(1) T},\vec{\lambda\,}_+^{(2) T},\vec{\lambda\,}_-^{(2) T})$, where
$\vec{\lambda\,}_{\pm}^{(n) T}$$=$$(\lambda_{\pm,1}^{(n)},\ldots, \lambda_{\pm,N}^{(n)})$, $n=1,2$. Moreover $S_{\pm}=(S_3\pm iS_2)/2$. Then
\begin{eqnarray}
H_{\rm I} &=& S_+ \otimes \vec{\Lambda\,}^T   \vec{A} + S_-\otimes \vec{A\,}^\dagger   \vec{\Lambda\,} + S_1\otimes: \vec{A\,}^\dagger   T  \vec{A\,}:
 \end{eqnarray}
The symbol $:ab:$ denotes normal ordering with respect to a thermal expectation value. The $4N\times 4N$ matrix $T$  has the following block structure
\begin{equation}
\label{Tmatrix}
T\ =\  \left(\begin{array}{cccc}
s_{11}&t_{11}& s_{12}&t_{12}\cr
t_{11}&s_{11}& t_{12}& s_{12}\cr	
s^T_{12}&t^T_{12}&s_{22}&t_{22}\cr
t^T_{12}&s^T_{12}&t_{22}& s_{22}
\end{array}\right) \ , \quad t_{ii}= t_{ii}^T,  \quad  s_{ii}= s_{ii}^T,\quad t_{ij}, s_{ij} \in {\mathbb R}\ .
\end{equation}
Note the invariance of $T$ under the unitary automorphism $T\to \Sigma_x^{-1} T  \Sigma_x $, where $\Sigma_x= {\mathbb 1}_2\otimes \sigma_x\otimes {\mathbb 1}_{N}$ 
and $\sigma_x=\left[\begin{array}{cc}0&1\cr 1&0\end{array}\right]$ is a Pauli matrix. Likewise we 
define $\Sigma_z$. The generator reads
\begin{eqnarray}
\eta &=& S_+\otimes \vec{\Lambda\,}^T  \left(\Delta- \Omega\right)   \vec{A\,} - S_- \otimes \vec{A\,}^\dagger   \left(\Delta- \Omega\right)   \vec{\Lambda\,} + 
                  S_1 \otimes : \vec{A\,}^\dagger   [\Omega,T]    \vec{A\,} : \ ,
\end{eqnarray}
where $\Omega$ $=$ $\diag(\omega^{(1)},-\omega^{(1)} ,\omega^{(2)},-\omega^{(2)})$ and $\omega^{(n)}$ $ =$ $\diag(\omega_1^{(n)},\ldots,\omega_N^{(n)})$, $n=1,2$. 
In former treatments of the Spin--Boson model with a single bath \cite{keh96, keh97} within the 
flow--equation approach, a formally simpler generator was used instead of the canonical one
$\eta= [H_0,H]$.  This reduced the number of differential equations to be solved. 
The different generators were contrasted in Ref.~\cite{sta03}. 

In general  there seem to exist by now no other guideline to 
improve the canonical generator than educated guess or physical intuition. Thus, for 
the present problem we stick to the canonical one.

The  commutator $[\eta,H]$ is calculated straightforwardly and a set of non--linear coupled ODE's is 
obtained for the tunnelling matrix element $\Delta$, the couplings $\vec{\Lambda}$ and for the matrix elements of $T$. They read
\begin{eqnarray}
\frac{d\Delta(B)}{dB}& = &\frac{1}{2} \vec{\Lambda\,}^T  \left(\Delta-\Omega\right)   \coth\left(\frac{\beta |\Omega|}{2}\right)  \vec{\Lambda\,}\ ,\nonumber\\
\frac{d\vec{\Lambda}(B)}{dB}& = &- \left(\Delta-\Omega\right)^2   \vec{\Lambda} + 
   \left\{T(\Delta-\Omega) + [\Omega,T]\right\}   \coth\left(\frac{\beta |\Omega|}{2}\right) \vec{\Lambda} \ ,\nonumber\\ 
\frac{d T (B)}{dB}& = &- [\Omega,[\Omega,T] ] - \frac{1}{2}\vec{\Lambda\,}\left(\Delta-\Omega\right)\vec{\Lambda\,}^T - \frac{1}{2}\Sigma_x\vec{\Lambda\,}\left(\Delta-\Omega\right)\vec{\Lambda\,}^T\Sigma_x\ .
                     \label{FloweqHam}               
\end{eqnarray}

The equations (\ref{FloweqHam}) form a set of $1+4N+2N(4N+1)$ first order non--linear differential equations which must be solved numerically. 
Before we do so, we show how they reduce to the RNG equations (\ref{ham}) for 
an ohmic bath in the low frequency limit. We limit ourselves to zero temperature. 
The differential equations for entries of $T$  are of the type
\begin{equation}
\frac{df(B)}{dB} \ =\ \omega f(B) + g(B)\ , \qquad \omega\in{\mathbb R}
\end{equation} 
which can be solved exactly
\begin{eqnarray}
f(B) & = & f(0)e^{\omega B} + \int_0^BdB^\prime e^{\omega(B-B^\prime)}g(B^\prime) \ .
\label{flowapprox}
\end{eqnarray}
This might be plugged into  the flow equation for $\vec{\Lambda}$. It suffices to evaluate these equations for small 
frequencies. Using the definitions of the spectral functions (\ref{spectralfunc}) and 
\begin{equation}
\sum_k \lambda_{j,k}^{(n)}(B)\lambda_{j^\prime,k}^{(n)}(B^\prime)\delta(\omega-\omega_k^{(n)}) \ = 
\ 2\sqrt{\gamma_j^{(n)}\gamma_{j^\prime}^{(n)}}\omega \ , \qquad \forall j,j^\prime\in \{2,3\}
\end{equation}
an integro--differential equation for the coupling constants is acquired
\begin{eqnarray}
\label{flowF2a}
\frac{d\gamma^{(n)}_{3}(B)}{dB}&=& -2 \Delta(B)^2 \gamma^{(n)}_{3} (B)-2 \int_0^B d B^\prime \sqrt
{\gamma^{(n)}_{3}(B) \gamma^{(n)}_{3}\left(B^\prime\right)} \omega_c^2\nonumber\\
                                                             &&\qquad\qquad \times  \int_0^{1}d x e^{-x\omega_c^2(B-B^
\prime)}\sum_{m=1}^2\left(\Delta(B)\sqrt{\gamma_{3}^{(m)}(B)}- 2\omega_c \sqrt{\gamma_{2}^{(m)}
(B) x} \right) \nonumber\\
                                                             &&\qquad\qquad\qquad\times\left(2\Delta(B^\prime)\sqrt
{\gamma_{3}^{(m)}(B^\prime)}- \omega_c \sqrt{\gamma_{2}^{(m)}(B^\prime) x} \right)  \ .                                                                  
\end{eqnarray}
The corresponding equation for $\gamma^{(n)}_{2}$ is obtained from Eq.~(\ref{flowF2a}) by 
interchanging the indices $2$ and $3$ everywhere. This equation allows for a perturbative 
expansion in $h=\Delta/\omega_c$. Keeping only the highest order 
term in the integral Eq.~(\ref{flowF2a}) reduces to
\begin{eqnarray}
\label{flowF2b}
\frac{d\gamma^{(n)}_{3}(B)}{dB}&=& -2 \omega_c^2 h^2(B) \gamma^{(n)}_{3} (B)-4 \int_0^B d B^
\prime \sqrt{\gamma^{(n)}_{3}(B) \gamma^{(n)}_{3}\left(B^\prime\right)} \omega_c^4 \nonumber\\
                                                             &&\qquad\qquad \times  \int_0^{1}d x x e^{-x\omega_c^2(B-B^
\prime)}\sum_{m=1}^2 \sqrt{\gamma_{2}^{(m)}(B)\gamma_{2}^{(m)}(B^\prime)  }  \ .                                                                  
\end{eqnarray}
In the limit $\omega_c\to \infty$ the $B^\prime$ integration becomes $\delta$--like for almost all $x
\in [0,1]$ and we arrive at
\begin{eqnarray}
\frac{d\gamma_3^{(n)}(B)}{dB} &=& -2 \omega_c^2 h^2(B) \gamma^{(n)}_{3} (B) - 4\gamma_3^{(n)}
(B)\omega_c^2 \sum_{m= 1}^2\gamma_2^{(m)}(B)    
\end{eqnarray}
and likewise for $\gamma_2^{(n)}(B)$. To make contact with the RNG equations, we use the 
relation \cite{keh06}
\begin{equation}
\omega_c \ = \ \frac{1}{2\sqrt{B}} \ =\ e^{-l} 
\end{equation}
and obtain 
\begin{eqnarray}
\label{rng1}
\frac{d\gamma_2^{(n)}(l)}{dl}&=&  h^2\gamma_2^{(n)} -2\gamma_2^{(n)} \sum_{m=1}^2 \gamma_3^{(m)}\nonumber\\
\frac{d\gamma_3^{(n)}(l)}{dl}&=&  h^2\gamma_3^{(n)} -2\gamma_3^{(n)} \sum_{m=1}^2 \gamma_2^{(m)}\ , \qquad n= 1, 2\ .
\end{eqnarray}
One derives straightforwardly from equation (\ref{FloweqHam})
\begin{equation}
\label{rng2}
\frac{d h(l)}{dl}\ = \  \left(1- \sum_{j= 2,3}\sum_{m=1}^2 \gamma_j^{(m)} \right) h\ .
\end{equation} 
Equations (\ref{rng1}) and (\ref{rng2}) correspond to the one--loop perturbative renormalization group equations for
 arbitrary couplings $\gamma_2^{(n)}$ and $\gamma_3^{(n)}$, $n=1,2$. We do not analyze them further here, but only mention that 
the result of Novais et al. \cite{nov05} stated in Eq.~(\ref{renequations}) is obtained by setting $
\gamma_2^{(1)}$ and $\gamma_3^{(2)}$ to zero. 
However it must be pointed out that the same equations are obtained for $\gamma_3^{(1)}$ and $\gamma_2^{(1)}$  , if $\gamma_3^{(2)}$ and $\gamma_2^{(2)}$ are set to zero, i.~e.~ in the absence of the second bath.

An adaptive step--size fourth order  Runge--Kutta algorithm has proved to be a reliable solver of
 the flow equations (\ref{FloweqHam}). Most entries of $\vec{\Lambda}$ and of $T$ become 
exponentially small for large flow parameter and the Hamiltonian becomes diagonal 
\begin{equation}
\label{Hinf}
H(\infty) \ =\ -\Delta_r S_1 +  \frac{1}{2}:\vec{A\,}^\dagger |\Omega |\vec{A\,}: + H_{\rm res}
\end{equation}
with a finite renormalized tunnelling matrix element $\Delta_r \equiv \Delta(\infty)$. Not all  entries of $\vec{\Lambda}$ and of $T$ 
decay exponentially for large $B$. From the flow equations (\ref{FloweqHam}) it is seen that the coupling matrix elements $\lambda_{+,k}^{(n)}$ for frequencies close 
to the renormalized tunnelling matrix element decay most slowly. On the other hand the diagonal entries of $T$ do not decay at all, leading 
 to an effective coupling of the bath modes to $S_1$ in the renormalized Hamiltonian 
 \begin{equation}
 H_{\rm res} = S_1\otimes \sum\limits_{n=1}^2 \sum_k s_{nn,kk}(\infty) a^\dagger_{n,k} a_{n,k} \ .
\end{equation}
Although this term -- being diagonal -- causes no additional difficulties, for practical purposes it can be neglected, since the residual matrix elements  $s_{11,kk}$, $s_{22,kk}$ 
are usually much smaller than the  mean level spacing of the bath modes.

\begin{figure}
\epsfig{figure=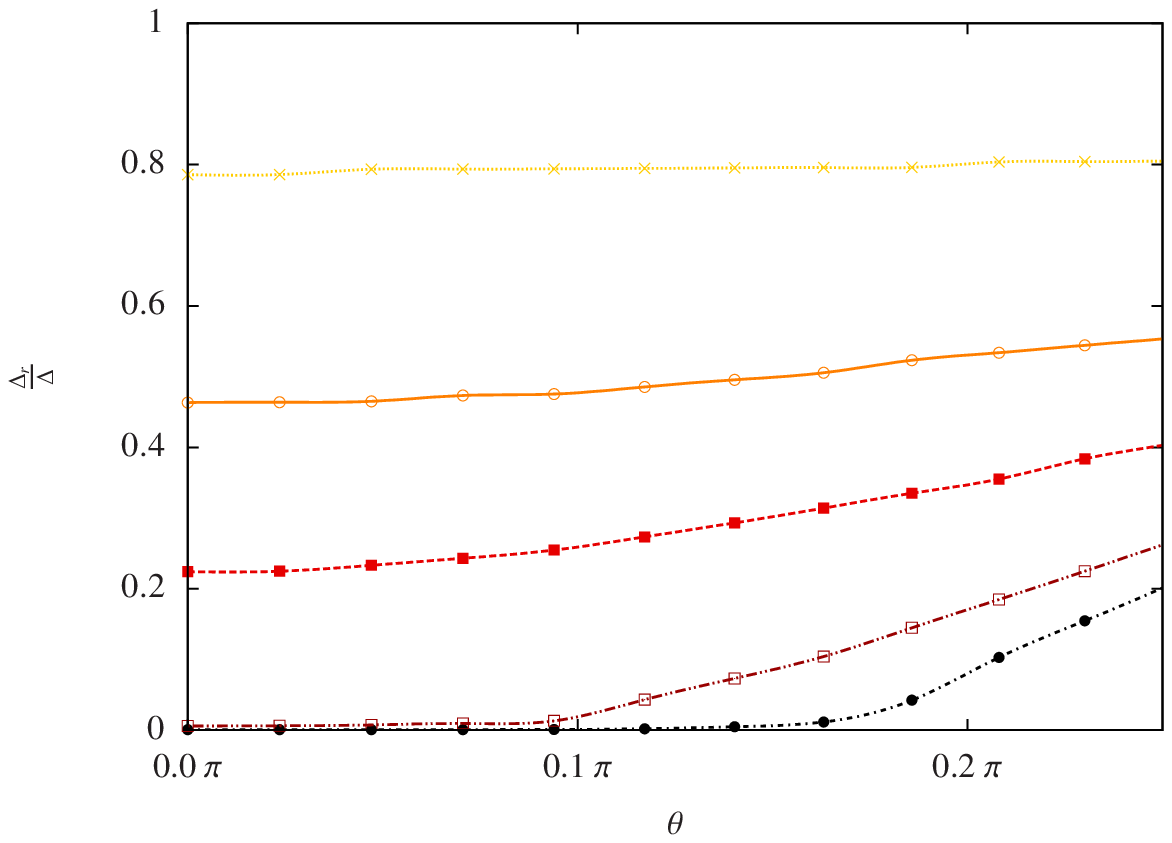, width= 8cm}
 \epsfig{figure=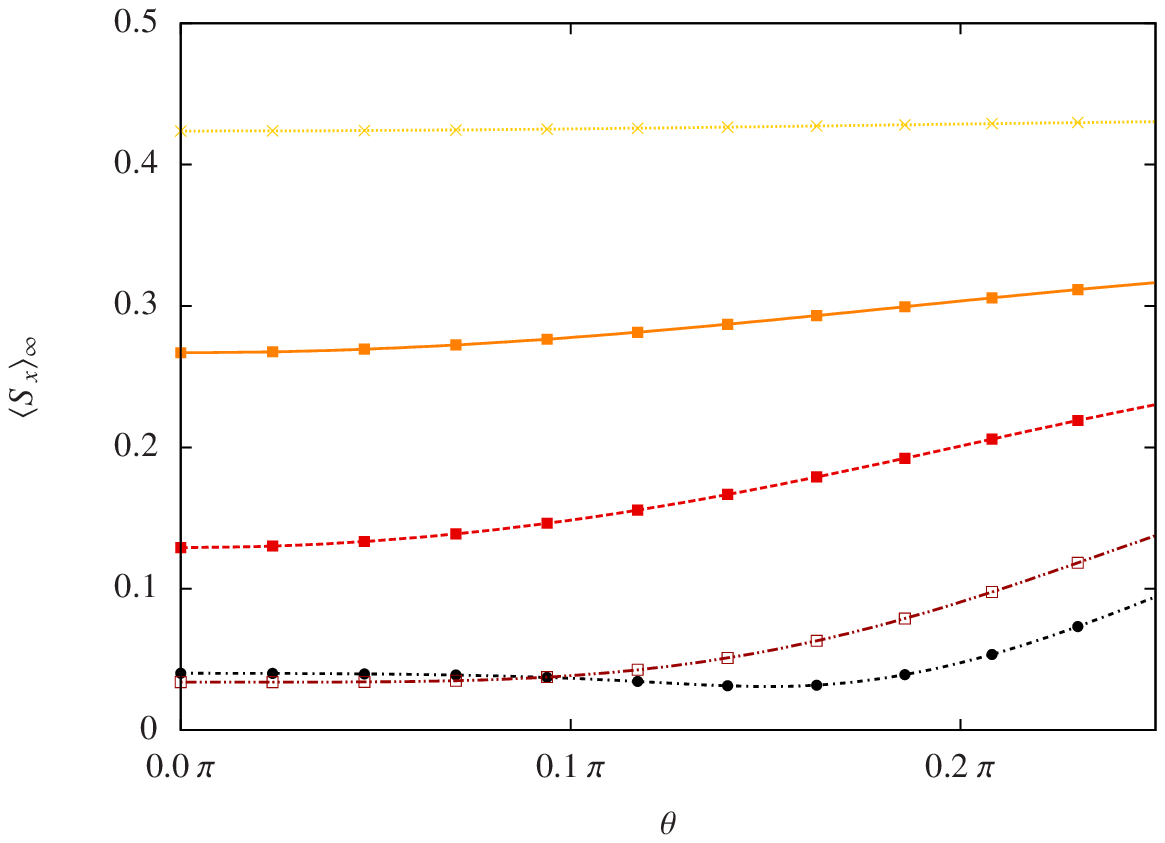,width=8cm}
\caption{\label{Fig1} Left: Plot of the renormalized tunneling matrix element $\Delta_r$ as a 
function of the angle $\theta$ defined in 
the main text, the total coupling strength is $\gamma_{\rm tot} = 0.1 $ (crosses, online yellow), $
\gamma_{\rm tot} = 0.3 $ (empty circles, online orange), $\gamma_{\rm tot} = 0.5 $ (filled boxes, 
online red),  
$\gamma_{\rm tot} = 0.8 $ (empty boxes, online dark red) and $\gamma_{\rm tot} = 1 $ (filled 
circles, full black line). The cutoff frequency is $\omega_c$ $=$ $10 \Delta$. The number of bath 
modes is $N=1000$.\\
Right: the same for the equilibrium expectation value $\langle S_1\rangle$. $\omega_c$ $=$ $10 
\Delta$, $N=400$.} 
\end{figure}

In figure \ref{Fig1}  the renormalized energy gap of the two--level system is plotted for a fixed 
overall coupling $\gamma_
{\rm tot}$ $\equiv$ $\gamma_{3}^{(1)} + \gamma_{2}^{(2)}$ as a function of the relative angle $
\theta
\equiv\arctan\left(\sqrt{\gamma_{2}^{(2)}/\gamma_{3}^{(1)} }\right)$ which varies from zero (single 
bath) to $\pi/4$ (equal coupling strength). Whereas for small overall coupling the 
renormalized energy gap $\Delta_r$ is almost independent of $\theta$, for increasing coupling 
strength the gap is protected by a symmetric  coupling. Finally for 
$\gamma_{\rm tot}$ $=1$ the energy gap renormalizes to zero for $\theta=0$ but remains finite 
for symmetric coupling. 

If  $\gamma_{\rm tot}$  is increased even further the energy gap $\Delta(B)$ crosses zero for some 
large value of $B$ and decays afterwards very slowly in an oscillatory fashion to zero. This 
happens for angles smaller than some critical angle, indicating the onset of the strong coupling 
regime, respectively of the KT phase transition.  
It is expected that the flow equations, being generically perturbative, become less exact for stronger 
coupling. However for $\theta = 0$  the critical value $\gamma=1$  was obtained analytically and 
with good precision numerically \cite{keh96}  . Therefore it is well justified to assume that the flow 
equations yield a good estimates  for the critical $\gamma_{\rm tot}$ for $\theta \neq 0$ as well.   

\begin{figure}
\epsfig{figure=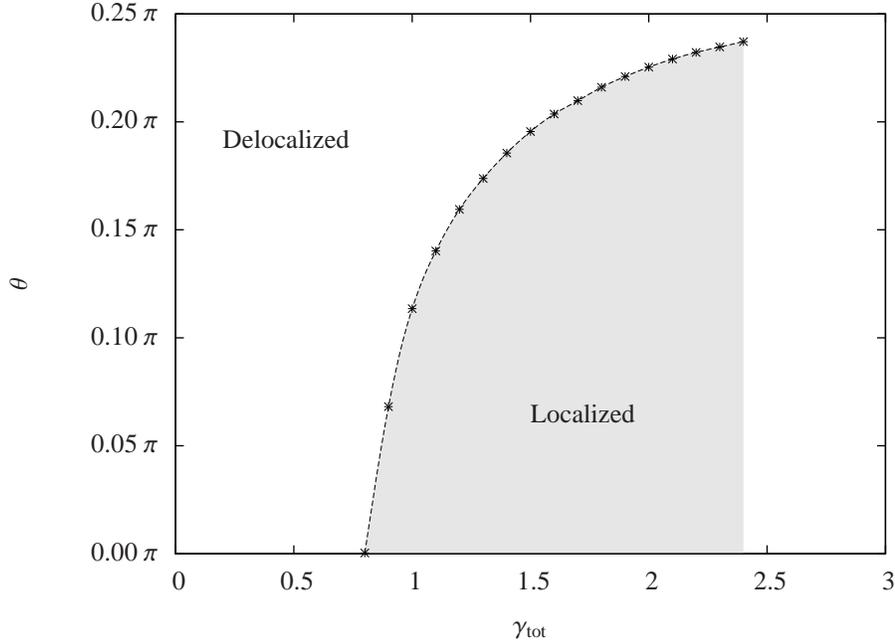}
\caption{\label{Fig5} Phase diagramm in the $\gamma_{\rm tot}$ -- $\theta$ plane. The line indicates the critical asymmetry angle, which separates the localized from the delocalized phase. The critical angle was determined for $N=800$ bath modes.} 
\end{figure}

In figure \ref{Fig5} the critical line is plotted  in the $\gamma_{\rm tot}$ -- $\theta$ plane, which 
separates the localized from the delocalized phase.  It is seen that it crosses the $x$--axis at some 
value smaller than one.  This offset is due to the finite number of bath modes and of the finite cutoff 
frequency. This  can  be improved systematically by increasing the number of bath modes and 
simultaneously increasing the endpoint of the flow $B_{\rm max}$. For values of $\gamma_{\rm tot}
$  larger a than some value $\gamma_{\rm tot}\approx 2.5$ the flow becomes unstable.

\section{Equilibrium expectation values}
In oder to calculate equilibrium expectation values with respect to the transformed Hamiltonian $H
(\infty)$ the corresponding operators have to transform as well.  Complex $4N$--vectors $\vec{\chi\,}$ and $\vec{\zeta\,}_{0,1}$ are 
introduced and the spin operators are expanded as  
\begin{eqnarray}
\label{ansatzred}
S_1 &=& h_0 S_0 + h_1S_1+ S_+ \otimes \vec{\chi\,}^{\dagger}  \vec{A}+ 
 S_- \otimes \vec{A\,}^\dagger    \vec{\chi\,}\nonumber\\
S_+ &=& h_+ S_+ + h_- S_- + S_0  \otimes \vec{\zeta\,}_0^{ \dagger}  \vec{A}+ 
		  S_1 \otimes \vec{\zeta\,}_1^{ \dagger}  \vec{A}\nonumber\\
S_- &=& h_-^* S_+ + h_+^* S_- +  S_0 \otimes  \vec{A\,}^\dagger   \vec{\zeta\,}_0 + 
 			S_1 \otimes \vec{A\,}^\dagger   \vec{\zeta\,}_1  
 \end{eqnarray}
The  flow equations for $h_0$, $h_1$, $h_\pm$ and for $\vec{\chi\,}$, $\vec{\zeta\,}_{0,1}$  are 
obtained by calculating the commutator $[\eta,S_i]$. The equations are closed by neglecting all normal 
ordered operator products with two or more annihilation or creation operators.  They are stated in  
App.~\ref{app2}.
The equilibrium density matrix with respect to the  renormalized free Hamiltonian (\ref{Hinf}) is just
\begin{equation}
\rho_{\rm eq}\ =\ \left( S_0 + \tanh\left(\frac{\Delta_r\beta}{2}\right)S_1\right)
                     \otimes \rho^{(1)}_{\rm eq}\otimes \rho^{(2)}_{\rm eq} \ .
\end{equation}
Here  $\rho^{(n)}_{\rm eq}$ $=$ $\prod_k \exp(-\beta \omega_k^{(n)} a_{n,k}^\dagger a_{n,k})/$ is the thermal  density matrix of the two free environments.
Thus, once the equations are numerically 
 solved, an arbitrary equilibrium expectation value of the spin operators is readily calculated. 
As an example we consider the one--sided Fourier transform 
\begin{equation}
\label{chiperp}
\chi_{zz}(\omega)\ =\ - i\int_0^\infty\frac{dt}{2\pi}e^{i\omega t}\left<[S_3(0),S_3(t)]\right>
\end{equation}
of the correlator $\left<[S_3(0),S_3(t)]\right>$ which was investigated in \cite{nov05}. At zero 
temperature, its imaginary part $\chi_{zz}^{\prime\prime}$ is given by  
\begin{eqnarray}
\chi^{\prime\prime}_{zz}(\omega)&\propto& \left(h_+ + h_-\right)^2 \delta\left(\omega-\Delta_r\right) +
\left(\vec{\zeta\,}_0 + \vec{\zeta\,}_1\right)^\dagger \left(1+\Sigma_x\right)\delta\left(\omega - |\Omega|\right) \left(\vec{\zeta\,}_0 + \vec{\zeta\,}_1\right) \ .
\end{eqnarray}
As a second example, we consider the equilibrium expectation value 
\begin{equation}
\langle S_1\rangle \ =\ \frac{h_0}{2} + \frac{h_1}{2} \tanh\left(\frac{\Delta_r\beta}{2}\right) \ .
\end{equation} 
 It is plotted in the bottom picture of figure \ref{Fig1}  for zero temperature and for different angles $\theta$ as defined before. Since the calculation is numerically more expensive than that of the energy gap, the number of bath modes is $N=400$. 
 For small and intermediate coupling it behaves qualitatively similar  to the renormalized two--level energy gap $\Delta_r$.  For strong coupling
 $\gamma_{\rm tot}$ $\approx$ $1$ it is seen that  $\langle S_1\rangle$ does not scale to zero for $\theta =0$ as expected, indicating that the flow equations lose accuracy in the strong coupling regime.     

Before we discuss the numerical results for the equilibrium correlation functions an explanatory remark is in order. 
A careful treatment of equilibrium correlation functions within the flow--equation approach requires high sophistication. For frequencies 
close to the renormalized tunnel matrix element $\Delta_r$ the flow converges only very slowly with $B_{\rm max}$, the endpoint of the numerical 
integration of the flow. Since the endpoint of the integration is itself limited by the density of the bath modes an accurate resolution would require an out of scale 
number of bath modes. As a consequence of this numerical limitation the equilibrium correlation functions have a two--peak structure: one broad maximum at a value
smaller than $\Delta_r$ and a second sharp peak right at $\Delta_r$, which is clearly unphysical.  

The problem can be overcome by employing constants of motion under the flow. This was done in \cite{keh97} for the one--bath spin Boson model. The result is a smooth curve 
with a single peak. But such constants of motion under the flow are not always easy to identify. 

We refrain from this procedure and show the curves for $\chi^{\prime\prime}_{zz}(\omega)$ obtained by fitting the numerical data with smoothing splines using an extremely high fidelity factor (of order $10^8$) everywhere but around $\Delta_r$, where it is quartically 
suppressed. 

\begin{figure}
\epsfig{figure=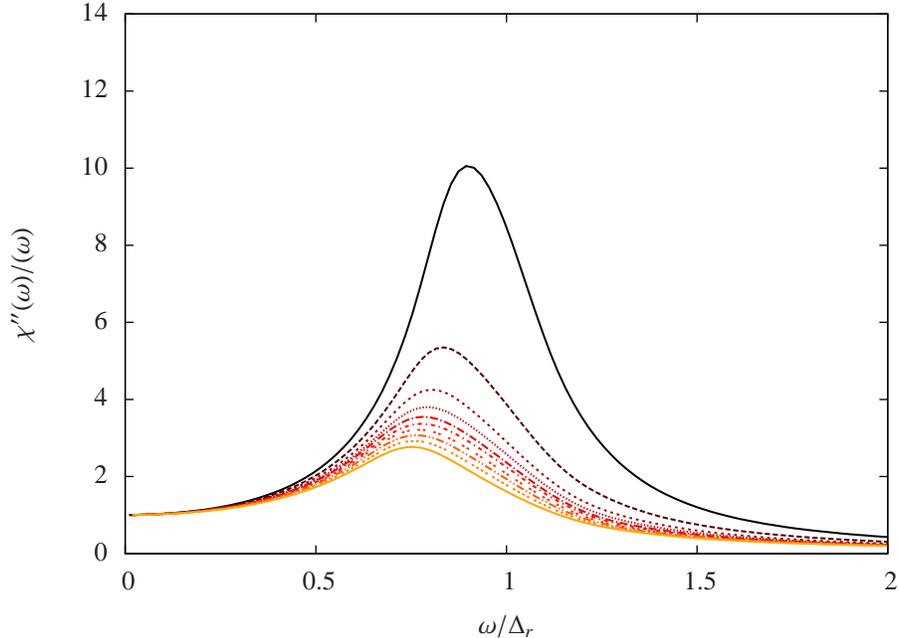}
\caption{\label{Fig2} Plot of the transverse susceptibility $\chi_{zz}^{\prime\prime}(\omega)/\omega$ in $z$--direction for symmetric coupling $\gamma_3^{(1)}$ $=$  $\gamma_2^{(2)}$ and for ten different values of $\gamma_{\rm tot}$ $= \sqrt{2} \cdot 0.1 \, n$, $1\leq n \leq 10$, from top to bottom (online color: from dark--colored to light--colored). The number of bath modes is 400,  $\Delta/\omega_c$ $=$ $1/10$.}
\end{figure}

In figure \ref{Fig2} the correlation function $\chi^{\prime\prime}_{zz}(\omega)$ is plotted for equal coupling strength to both baths and with an overall coupling strength $\gamma_{\rm tot}$ varying between $0.1$ and one. 
The curve corresponds to Fig.~4 in reference \cite{nov05} and is qualitatively similar. As the coupling strength increases the resonance peak becomes smaller and smaller but never disappears.  The maximum of the resonance peak is systematically 
below $\Delta_r$. This is a difference to Fig.~4 in reference \cite{nov05} where the maximum seems to be always right at the renormalized tunnel matrix element.

In figure \ref{Fig2} the correlation function $\chi^{\prime\prime}_{zz}(\omega)$ is plotted for fixed overall coupling strength $\gamma_{\rm tot}$ and for different angles $\theta$. The resonance peak in the symmetric case ($\theta=\pi/4$) is largely enhanced as compared to the highly asymmetric case ($\theta= 0.1 \pi$). However, the reason for this is rather trivial.  In the highly asymmetric case the coupling to the $z$--component is largest, whereas there is no coupling to the $y$--component. In the symmetric case the coupling to the $z$--component is reduced, which is reflected by the enhanced resonance peak of $\chi^{\prime\prime}_{zz}$. However the coupling to the $y$--component is larger, which yields a reduced  resonance peak of $\chi^{\prime\prime}_{yy}$ (not shown here).  If we write $\chi^{\prime\prime}_{zz}(\omega,\theta)$ as a function of the relative angle $\theta$, then the obvious relation  $\chi^{\prime\prime}_{zz}(\omega,\theta)$ $=$ $\chi^{\prime\prime}_{yy}(\omega,\pi/2-\theta)$ holds. Thus an enhancement of the resonance peak in $z$--direction comes necessarily with a decrease in $y$--direction and vice versa. Indeed in Fig.~\ref{Fig2} the resonance peak of $\chi^{\prime\prime}_{zz}$ is biggest for $\theta= 0.3\pi$ in spite of the asymmetric coupling (for even higher $\theta$ it increases more and more). Note however that the location of the maximum of the peak is maximal in the symmetric case. 
\begin{figure}
\epsfig{figure=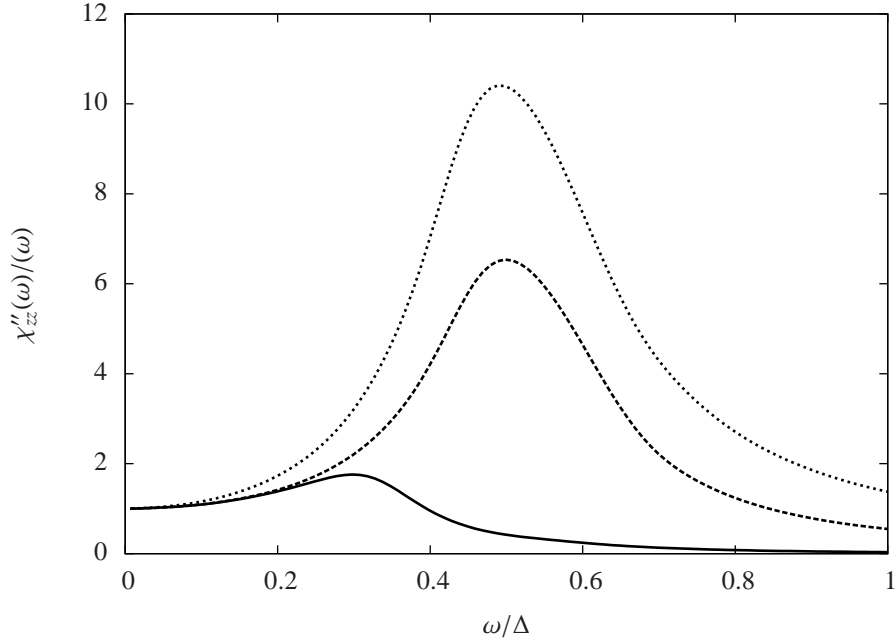}
\caption{\label{Fig3} Plot of the transverse susceptibility $\chi_{zz}^{\prime\prime}(\omega)/\omega$ in $z$--direction for three different angles $\theta$ $= 0.1\pi$ (full line), $\theta$ $= \pi/4$ (dashed line) and $\theta$ $= 0.3\pi $ (dotted line)  for overall coupling strength  $\gamma_{\rm tot} = 0.3$. The number of bath modes is $N=400$, $\omega_c/\Delta =$$10$.}
\end{figure}


\section{Thermalization and Decoherence}
\label{therm}

In thermal equilibrium the mutual energy transfer from the system to the environment and vice 
versa is zero, warranted by fluctuation dissipation theorems. However in the process of 
thermalization the net energy transfer of the system to the environment is positive.  

Assuming an decoupled initial state, which is fully polarized in some direction perpendicular to the $x$ axis (we may assuume $0\leq \theta^\prime \leq \pi/4$) 
\begin{equation}
\label{noneqdens}
\rho_{\rm init} \ = \ \left(S_0 + \cos\theta^\prime S_3+ \sin\theta^\prime S_2 \right)
                     \otimes \rho^{(1)}_{\rm eq}\otimes \rho^{(2)}_{\rm eq} \ ,
\end{equation}
thermalization is characterized by the time evolution of the expectation value of the system's 
energy $\langle H_{S}(t)\rangle$ $=$ $-\Delta \langle S_1(t)\rangle$ . 
This quantity is expected to approach its equilibrium value on a certain time scale, the so called 
relaxation time, which is usually denoted $T_1$. 

Decoherence is the creation of entanglement of the system with the environment. It is measured by 
the decay of the off--diagonal elements of the reduced density matrix 
of the spin in the $S_1$ basis, i.~e.~by the expectation values $\langle S_\pm\rangle$. A basis 
independent measure for decoherence is the 
purity ${\cal P}(t)$ $= $  ${\cal P}_\parallel(t)$ $+{\cal P}_\perp(t)$, 
where ${\cal P}_\parallel(t)$ $= $ 
$2 \sum_{n=0}^1 \langle S_n(t)\rangle^2$ and   ${\cal P}_\perp(t)$ $=$  $2 \sum_{n=2}^3 \langle 
S_n(t)\rangle^2$. Decay of decoherence takes place on a time scale $T_2$, called decoherence 
time \cite{sli96}, we asociate it with ${\cal P}_\perp(t)$. Both decoherence time and relaxation time enter in the definition of purity. We call the two quantities ${\cal P}_\perp(t)$ and ${\cal P}_\parallel(t)$ transverse 
respectively parallel purity. For the initial state (\ref{noneqdens}) ${\cal P}_\perp(0)$ $={\cal P}_\parallel(0)$ $=1/2$.

Assuming a decoupled initial state as in Eq.~(\ref{noneqdens}) first order differential equations for the spin expectation values are straightforwardly derived in second order perturbation theory 
\begin{eqnarray}
\label{bloch1}
\frac{d}{d t}\langle S_1\rangle &=& -\left( \Gamma^{(2)}_2(t)+ \Gamma^{(1)}_3(t)\right) \langle S_1\rangle - F(t)\nonumber\\
\frac{d}{d t}\langle S_2\rangle &=& \tilde{\Delta}^{(1)}_3(t)\langle S_3\rangle -  \Gamma^{(1)}_3(t) \langle S_2\rangle \nonumber\\
\frac{d}{d t}\langle S_3\rangle &=& -\tilde{\Delta}^{(2)}_2(t) \langle S_2\rangle -  \Gamma^{(2)}_2(t) \langle S_3\rangle \ ,
\end{eqnarray}
with the time--dependent coefficients
\begin{eqnarray}
 \Gamma^{(m)}_n(t) &=& \int\limits_0^t dt^\prime \int\limits_0^\infty d\omega  \cos\left(\Delta( t^\prime - t)\right)\cos\left(\omega(t-t^\prime)\right) J^{(m)}_n(\omega)\coth\left(\omega\beta/2\right) \nonumber\\
  \tilde{\Delta}^{(m)}_n(t) &=& \Delta- \int\limits_0^t dt^\prime \int\limits_0^\infty d\omega  \sin\left(\Delta( t^\prime - t)\right)\cos\left(\omega(t-t^\prime)\right) J^{(m)}_n(\omega)\coth\left(\omega\beta/2\right) \nonumber\\
F(t)&=& \int\limits_0^t dt^\prime \int\limits_0^\infty d\omega \sin\left(\omega (t-t^\prime)\right) \sin\left(\Delta (t^\prime-t)\right)  \left(J^{(2)}_2(\omega)+J^{(1)}_3(\omega)\right) \ .
\end{eqnarray}
In the Markov approximation these coefficients become time independent $\Gamma^{(m)}_n(t) = \Gamma^{(m)}_n$ $ = (\pi/2) J^{(m)}_n(\Delta)\coth(\beta \Delta/2)$, $F(t)= F= (\pi/2) (J^{(2)}_2(\Delta)+J^{(1)}_3(\Delta)) $ and 
$\tilde{\Delta}^{(m)}_n(t)=$ $\tilde{\Delta}^{(m)}_n =$
$\Delta- \Delta\int_0^{\omega_c} d\omega$ $ \coth(\beta \omega/2)$ $J^{(m)}_n(\omega)/(\omega^2-\Delta^2)$.  Note that for an ohmic bath and at zero temperature  $\tilde{\Delta}^{(m)}_n$ has a logarithmic singularity in the cutoff frequency $\omega_c$.  

From equations (\ref{bloch1}) the phenomenological Bloch equations are obtained
which predict an exponential decay of decoherence and of relaxation.  Their solutions are
\begin{eqnarray}
\label{blochsol}
\langle S_1(t)\rangle &=& (\langle S_1(0)\rangle- \langle S_1\rangle_{\rm eq}) e^{-(\Gamma^{(2)}_2+\Gamma^{(1)}_3)t} + \langle S_1\rangle_{\rm eq}\nonumber\\
\langle S_{n}(t)\rangle &=& \frac{\lambda_+\langle S_n(0)\rangle+i\langle\dot{S}_n(0)\rangle}{\lambda_+-\lambda_-} e^{i\lambda_- t} -  \frac{\lambda_-\langle S_n(0)\rangle+i\langle\dot{S}_n(0)\rangle}{\lambda_+-\lambda_-} e^{i\lambda_+ t}
\end{eqnarray} 
where $\lambda_\pm$  are the roots of the characteristic polynomial
\begin{equation}
\chi(\omega) \ =\ \omega^2-i\omega(\Gamma^{(2)}_2+\Gamma^{(1)}_3)+ \tilde{\Delta}^{(2)}_2\tilde{\Delta}^{(1)}_3 -\Gamma^{(2)}_2\Gamma^{(1)}_3
\end{equation}
Thus decoherence and relaxation time are given by $T_1=1/(\Gamma^{(2)}_2+\Gamma^{(1)}_3)$ and $T_2$$=$ $2T_1$.  In second order perturbation theory the friction coefficients of the two baths add up. No frustration occurs.

In the Markov approximation Bloch equations hold beyond perturbation theory with relaxation and decoherence times depending in a more complicated non--perturbative way on the coupling strength $\gamma^{(1)}_3$ and $\gamma^{(2)}_2$. Corrections were calculated in Ref.~\cite{nov05}. In the regime where the Bloch equations (\ref{bloch1}) hold, the quantum regression theorem can be invoked and the dynamics of the expectation values is governed by the equilibrium  correlation functions.   
 
Yet at low temperature and on the time scale of the inverse cutoff frequency,  Bloch equations do not hold.  The coefficients in Eq.~(\ref{bloch1}) become time dependent and the simple exponential behavior (\ref{blochsol}) breaks down. This is seen most directly 
in a Taylor expansion of the time--evolution operator $U(t)= 1- iH t- t^2 H^2/2+ {\cal O}(t^3)$. For the initial state (\ref{noneqdens}) it predicts a quadratic behavior of $\langle S_1\rangle = $ $t^2/2 \tau^2+ {\cal O}(t^3)$, where 
$ \tau^{-1}$ $\approx $ $\omega_c\sqrt{\gamma^{(1)}_3+\gamma^{(2)}_2}$ is the inverse quantum Zeno time. For the transverse purity one obtains
\begin{equation}
\label{shorttime}
2P_{\perp}(t) \ =\ 1- \frac{t^2}{\tau^2}\left\{\sin^2(\theta)\cos^2(\theta^\prime)+\sin^2(\theta)\cos^2(\theta^\prime)\right\}+ {\cal O}(t^3)
\end{equation} 
The quadratic time dependence vanishes iff $\theta = \theta^\prime$ $= 0$. This indicates that initially, for short times, a symmetric coupling accelerates decay of coherence.

 In order to monitor the time evolution of the expectation values in the transient regime on a time scale of order of the quantum Zeno time, methods of non--equilibrium real time thermodynamics must be employed.     
Real time quantum evolution is addressed within the flow equation approach \cite{hack08,hack09} 
by applying subsequently the unitary transformation $U_B(B_1,B_2)$ generated by $\eta(B)$ and 
the time evolution operator $U_{t,\infty} (t_1,t_2)= e^{-iH(\infty)(t_1-t_2)}$ on the operator of interest 
according to the diagramm:
\begin{center}
\begin{picture}(200,120)
\put(0,100){${\cal O}(B=0,t=0)$}
\put(100,100){\vector(1,0){50}}
\put(102,110){$U_B(0,\infty)$}
\put(160,100){${\cal O}(\infty,0)$}
\put(180,90){\vector(0,-1){50}}
\put(182,60){$U_{t,\infty}(0,t)$}
\put(160,20){${\cal O}(\infty,t)$}
\put(150,20){\vector(-1,0){50}}
\put(102,30){$U_B(0,-\infty)$}
\put(0,20){${\cal O}(0 ,t)$}
\put(20,90){\vector(0,-1){50}}
\end{picture}
\end{center}
Since time evolution is simple for $B=\infty$  the observables are first transformed into the $B=\infty$ 
basis evolve in time and are then transformed back. At time $t$ the Heisenberg operators 
have been propagated by the diagonalized Hamiltonian (\ref{Hinf}). This yields new time dependent expansion coefficients $\tilde{h\,}_{\pm}(t)$$=$ $e^{\pm i \Delta_r t}h_\pm(\infty)$, $\vec{\tilde{\chi}\,}(t) $ $= $ $e^{i(\Delta_r-\Omega)t} $ $\vec{\chi}(\infty)$ and
$\vec{\tilde{\zeta}\,}_n(t)$ $ = $ $e^{i\Omega t} \vec{\zeta}_n(\infty)$, $n=0,1$. The coefficients $h_0=\tilde{h\,}_0$ and $h_1=\tilde{h\,}_1$ remain constant under time evolution. 
These coefficients are numerically transformed back, yielding an approximate solution of 
the Heisenberg equation for the spin operators. The expectation value with respect to the density matrix (\ref{noneqdens}) are 
\begin{eqnarray}
\langle S_1(t)\rangle & = & \tilde{h}_0 (t)/2\nonumber\\ 
\langle S_2(t)\rangle  & = & {\rm Im}\left[(\tilde{h}_+(t)-\tilde{h}^*_- (t))e^{i \theta^\prime}/2\right]\nonumber\\  
\langle S_3(t)\rangle &= &{\rm Re}\left[(\tilde{h}_+(t)+\tilde{h}^*_- (t))e^{i \theta^\prime}/2\right]. 
\end{eqnarray}
The calculation is numerically delicate \cite{hack08,hack09}. In order to perform the backward integration the forward flow of the Hamiltonian must be stored. This is a sizable 
amount of data of order of one terabyte. The read--in and the read--out slow down the routine. We thus performed the calculation of ${\cal P}_{\perp}$ with 250 bath modes, respectively of $\langle S_x(t)\rangle$ with 100 bath modes.     
\begin{figure}
\epsfig{figure=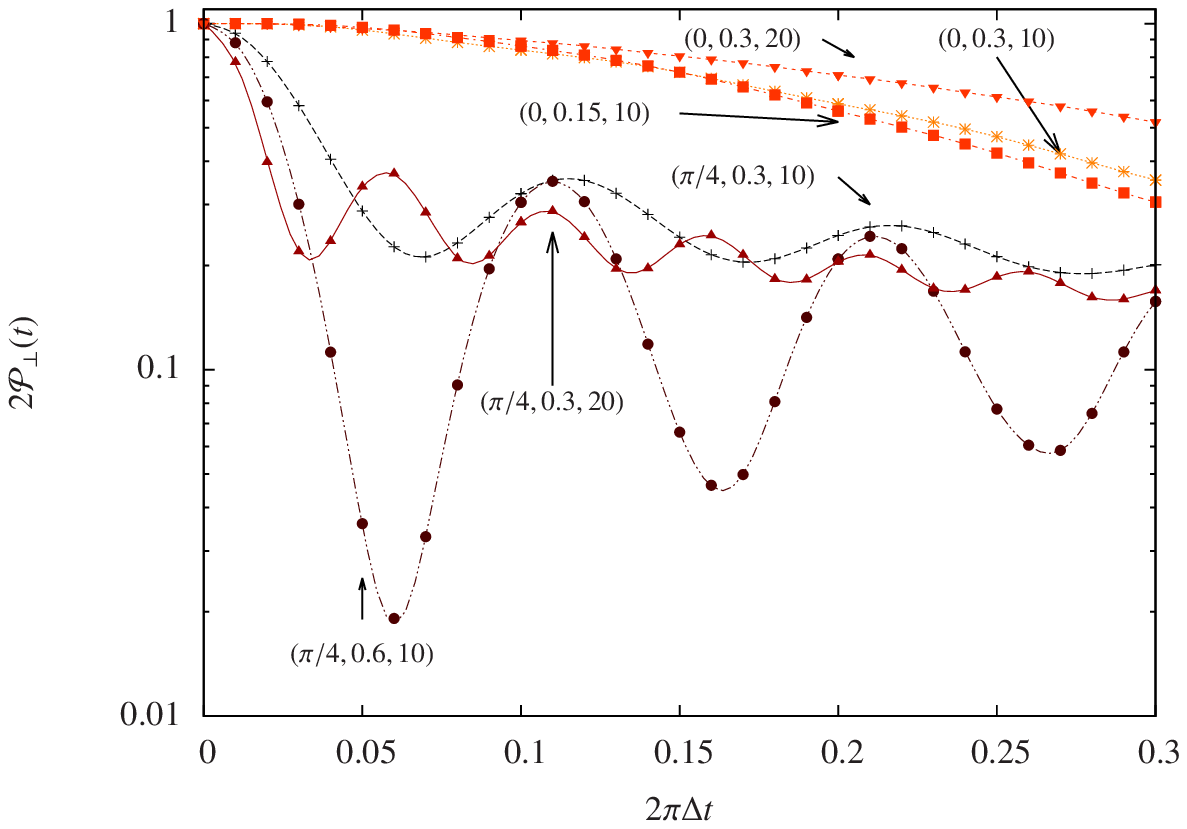,width=8cm}
\epsfig{figure=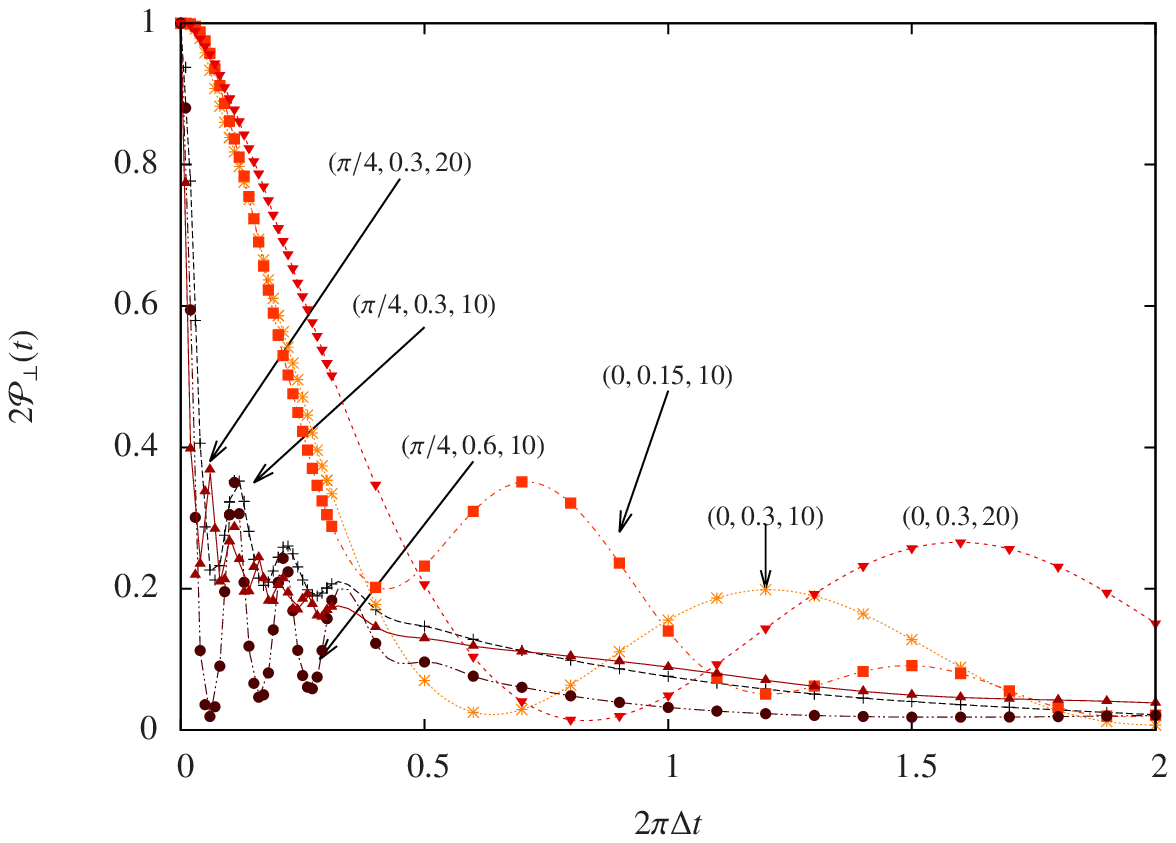,width=8.cm}
\caption{\label{Fig6} 
Right: Time evolution of  the transversal purity for an initial state characterized by the angle $\theta^\prime=0$ for different triplets $(\theta,\gamma_{\rm tot},\omega_c/\Delta)$. These are  $(\pi/4,0.3,10)$ (online black, crosses), $(0,0.3,10)$ (online 
yellow, asterisks), $(0,0.15,10)$ (online light red, boxes), $(\pi/4,0.6,10)$ (online dark purple, dots), $(\pi/4,0.3,20)$ (online lighter purple, triangles), $(0,0.3,20)$ (online darker red, triangles).\\
Left: The same for short times on a logscale.}
\end{figure}

In figure \ref{Fig6} the transverse purity is plotted for different values of $\gamma_3^{(1)}$ and  $\gamma_2^{(2)}$ for an initial state characterized by the angle $\theta^\prime = 0$. It is seen that for short times of order of $\omega_c^{-1}$ the transverse purity decays 
faster for a symmetric coupling than for a single bath, as predicted by equation (\ref{shorttime}). The decay occurs in an oscillatory fashion for both a single bath and for symmetric ccoupling. Although the dissipative two--level system has been studied extensively \cite
{leg87} to our best knowledge this oscillatory purity revival was not reported before. By now we do not have a satisfactory physical explanation for it. For symmetric coupling the oscillations decrease rapidly in less than one period of the Rabi oscillations. As can be 
seen from left picture of Fig.~\ref{Fig6} the  frequency seems to scale with $\omega_c$ and the amplitudes with $\gamma_{\rm tot}$. For a single bath the oscillations are much slower and decay less rapidly. 
\begin{figure}
\epsfig{figure=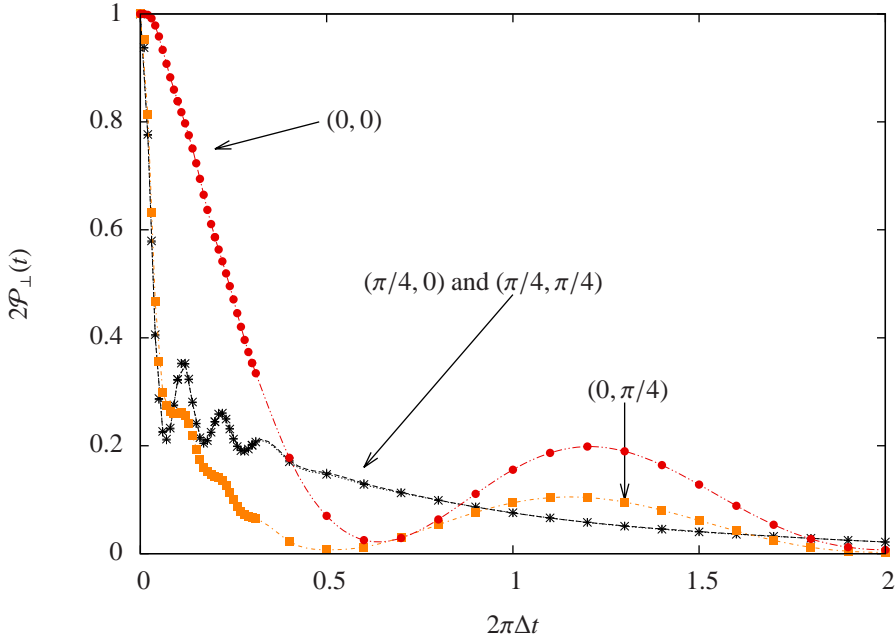}
\caption{\label{Fig6b} Time evolution of  the transversal purity for an initial state characterized by the angles $\theta^\prime=0,\pi/4$ for symmetric coupling (asterisks, online black) and for a single bath, $\theta^\prime$ $= 0$ (dotted, online red) and  $\theta^\prime$ $= \pi/4$ (boxed, online orange). The other parameters  are $\gamma_{\rm tot}=0.3$ and $\omega_c/\Delta$ $=10$.}
\end{figure}

The dependence on the initial state is considered in Fig.~\ref{Fig6b}. The transverse purity  for the initial state characterized by $\theta^\prime=0$ and 
for the initial state  $\theta^\prime=\pi/4$ is plotted. Whereas for symmetric coupling there is no visible difference, for a single bath the initial 
decay is much faster for $\theta^\prime = \pi/4$ than for $\theta^\prime = 0$, see Eq.~(\ref{shorttime}).  

\begin{figure}
\epsfig{figure=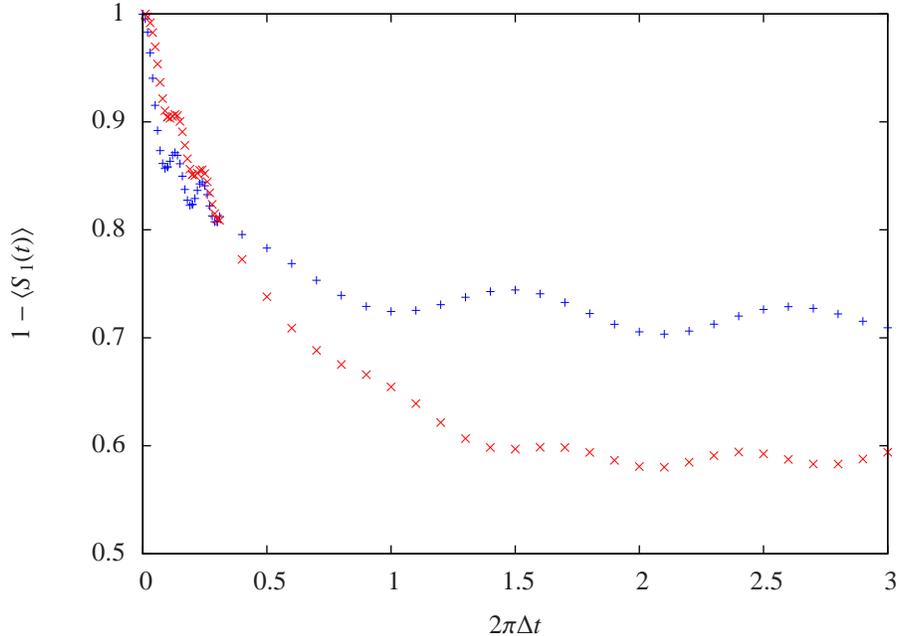}
\caption{\label{Fig7} Time evolution of  the expectation value $\langle S_x(t)\rangle$ for the initial state (\ref{noneqdens}) for symmetric coupling (online blue, stars) and for a single bath (online red, crosses), $\gamma_{\rm tot}=0.3$, $\omega_c/\Delta$$=10$. }
\end{figure}

The time evolution of $\langle S_x(t)\rangle$ is plotted in Fig.~\ref{Fig7} for symmetric coupling ($\theta=\pi/4$) and for a single bath ($\theta=\pi/4$)  
for a moderate overall coupling strength $\gamma_{\rm tot}=0.3$. Here the expectation value indeed decays initially faster for a single bath than for symmetric coupling. However on a time scale of the Rabi--oscillations the decay grows faster for symmetric coupling  to reach an equilibrium value, which is smaller than for a single bath in accordance with Fig.~\ref{Fig1}.

\section{Summary \& Discussion}
While the calculation of equilibrium correlation functions is somewhat cumbersome within the flow equation approach, the method turns out to be a useful numerical tool in non--equilibrium physics.  We were able to monitor purity decay on the time scale of the quantum Zeno time as well as on the time scale of the inverse Rabi frequency.  

When one speaks about coherence of a two--level system one has carefully to distinguish between 
the decay of the off--diagonal elements and of the diagonal elements. It is characteristic for a small 
size Hilbert--space that both are not independent and the distinction between decoherence and 
dissipation is fuzzy.  

In our analysis frustration effects of two independent oscillator bath could be identified in the 
renormalized energy gap $\Delta_r$, in the ground state expectation value of $S_1$ and in the 
ground state energy shift. These quantities are protected by a symmetric coupling. In particular  the 
protection of $\langle S_1\rangle$ can rightly be called protection of decoherence since it 
contributes to a high equilibrium purity of the spin.   

In non--equilibrium relaxation, i.~e.~the decay of $\langle H_S\rangle$ $\propto$ $-\langle 
S_1\rangle$, is protected by a symmetric coupling on a time scale of the quantum Zeno time. 
However, the decay of the off--diagonal matrix elements  of the reduced density matrix, corresponding to $\langle S_2\rangle$, $\langle S_3\rangle$ and to the transverse purity is systematically faster for a symmetric coupling.   

The decay of both  $\langle H_S\rangle$ and of the transverse purity occurs in an oscillatory fashion. The physical reason behind these oscillations is unclear.

The dependence of the renormalized energy gap $\Delta_r$ on an asymmetry angle is a generic 
non perturbative effect. The flow equations (\ref{FloweqHam}) might be truncated by setting all 
second order terms, ~i.~e.~the matrix entries of $T$ (Eq.~(\ref{Tmatrix})), to zero. The truncated 
flow equations can be analyzed analytically, see App.~\ref{AppA}. The outcome is  
$\ln (\Delta_r/\Delta)$ $\propto -\gamma_{\rm tot}/(1- \gamma_{\rm tot})$,  
similar to the old result by Silbey and Harris \cite{sil83} which features no dependence on the asymmetry angle. Our analysis affirms that the delocalized phase for couplings $1< \gamma_{\rm tot}<\infty$ is stable against small asymmetries. 

The perturbative RNG equations  (\ref{rng1}) and  (\ref{rng2}) obtained from the flow equations are completely symmetric in the four coupling constants $\gamma_2^{(n)}$, $\gamma_3^{(n)}$, $n=1,2$. Setting any two of them to zero yields the RNG equations of Ref.~\cite{cas03}, with the implication of a delocalized phase for  $\gamma_{\rm tot}$ $\to\infty$. Setting for instance  
$\gamma_2^{(2)}=$$\gamma_3^{(2)}$ $=0$, this implys that also a symmetric coupling of the spin with its $y$ and $z$ components to a single bath can protect the delocalized phase. This question requires further investigation.

\acknowledgments
HK acknowledges financial support from the German Research 
council (DFG)  with grant No.~Ko 3538/1-2 and from CSIC within the JAE-Doc program  cofunded by the  FSE (Fondo Social Europeo) .  
AH Acknowledges support by the David and Ellen Lee foundation.
We acknowledge useful discussions with 
F. Guinea, F. Sols and T. Stauber. The computer cluster of the University of Duisburg--Essen was used for the numerics.

\appendix

\section{Linearized Flow equations for two baths}
\label{AppA}

We consider the linearized version of the flow equations. In the linearized version of the flow 
equations the flow of $T$ can be neglected.
\begin{eqnarray}
\frac{d\vec{\Lambda\,}(B)}{dB} &=& -\left(\Delta -\Omega \right)\vec{\Lambda}
\end{eqnarray}
For ohmic spectral functions $J_{i}^{(n)}(\omega)$ $=$ $ 2 \gamma_{i}^{(n)}\omega\theta
(\omega_c-\omega)$ , $i=2,3$, $n=1,2$
immediately the first order RG equations
\begin{eqnarray}
\frac{d\gamma_{i}^{(n)}}{dB} &=& - \Delta^2 \gamma_{i}^{(n)}\qquad i=2,3,\ \quad n=1,2
\end{eqnarray}
are obtained. Introducing the auxiliary densities
\begin{equation}
G_{\pm}^{(n)}(\omega)\ =\  \sum_{k}\left(\lambda_{\pm,k} ^{(n)}\right)^2
\delta\left(\omega-\omega^{(n)}_k\right)\qquad n=1,2
\end{equation}
 the renormalization group equation for the tunneling matrix element (\ref{FloweqHam}) can be 
written as
\begin{eqnarray}
\frac{d\Delta(B)}{dB} &=& -\frac{1}{4}\sum_{n=1}^2\sum_{\sigma=\pm} \int d\omega \frac{\coth(\beta\omega/2) }{\Delta(B)- \sigma\omega} \frac{d}{d B} G_{\sigma}^{(n)}(\omega,B)  
\end{eqnarray}   
Following the outlines of \cite{keh06} a self consistency equation for $\Delta_r$ can be obtained. 
For zero temperature it reads
\begin{eqnarray}
\ln \frac{\Delta_r}{\Delta} &=& \sum_{n=1}^2\sum_{\sigma=\pm} \int\limits_0^\infty 
                \frac{d\omega }{4\Delta_r}\frac{G_{\sigma}^{(n)}(\omega,0)}{\Delta_r -\sigma\omega}
\end{eqnarray}
For $\lambda^{(1)}_{2,k}$ $=$ $\lambda_{3,k}^{(2)}$ $=0$, $G_{+}^{(1)}$ $=G_{-}^{(1)}$ $= J_{3}^
{(1)}$ and $G_{+}^{(2)}$ $=G_{-}^{(2)}$ $= J_{2}^{(2)}$ and for an ohmic bath the renormalized 
matrix element becomes
\begin{equation}
\Delta_r\ =\  \Delta \left(\frac{\Delta}{\omega_c}\right)^{\frac{\gamma_3^{(1)}+ \gamma_2^{(2)}}{1-
\gamma_2^{(1)}- \gamma_3^{(2)}}}\ .
\end{equation}
This  is a straightforward extension of the old result by Silbey and Harris \cite{sil83}. 
In the linear approximation of the flow equations there is no angle dependence of $\Delta_r$.  The full flow 
equations must be employed.

\section{Flow equations for the spin operators}
\label{app2}
The flow equations for the expansion coefficients of the spin--operators are obtained from the commutators $[\eta,S_1]$ and $[\eta,S_\pm]$. They read:
\begin{eqnarray}
\label{appB1}
\frac{d h_0(B)}{dB} &=& \frac{1}{2}\vec{\Lambda\,}^T(\Delta-\Omega)\Sigma_z\vec{\chi\,} \nonumber \\
\frac{d h_1(B)}{dB} &=& -\frac{1}{2}\vec{\Lambda\,}^T(\Delta-\Omega)\coth\left(\frac{\beta|\Omega|}
 {2}\right)\vec{\chi\,}\nonumber\\
\frac{d h_+(B)}{dB} &=& \frac{1}{2}\left(\vec{\zeta\,}_1^\dagger \coth\left(\frac{\beta|\Omega|} {2}\right)- \vec{\zeta\,}_0^\dagger \Sigma_z\right)\, \Sigma_x\,(\Delta-\Omega)\vec{\Lambda\,}\nonumber\\
\frac{d h_-(B)}{dB} &=&   \frac{1}{2}\left(\vec{\zeta\,}_1^\dagger \coth\left(\frac{\beta|\Omega|} {2}\right)+ \vec{\zeta\,}_0^\dagger \Sigma_z\right)\, (\Delta-\Omega)\vec{\Lambda\,}\nonumber\\
\frac{d \vec{\chi}(B)}{dB}&=& h_1(\Delta-\Omega)\vec{\Lambda}
+[\Omega,T]\coth\left(\frac{\beta|\Omega|}{2}\right)\vec{\chi\,} \nonumber\\
\frac{d \vec{\zeta}_0(B)}{dB}&=&+[\Omega,T]\Sigma_z\vec{\zeta\,}_1
\nonumber\\
\frac{d \vec{\zeta}_1(B)}{dB}&=&-\frac{1}{2}\left(h_-^*  +h_+^* \Sigma_x \right) (\Delta-\Omega)\vec{\Lambda\,}+[\Omega,T]\Sigma_z \vec{\zeta\,}_0
\end{eqnarray}
These differential equations are the same for the forward flow and for the backward flow. However the initial conditions are 
different. For the forward flow the initial conditions are $h_1$ $=$ $h_+$ $=$ $1$ and all other components are zero. Since the differential equations are linear in the expansion coefficients the 
imaginary parts of $h_\pm$, $\vec{\chi}$ and $\vec{\zeta\,}_{0,1}$ remain zero throughout the flow. 

 Due to the time evolution the imaginary parts  acquire a non--trivial  backward flow.  The initial conditions are now 
 ${\rm Re\,} \vec{\tilde{\chi}\,}(t,0)$ $ =$ $  \cos((\Omega -\Delta_r) t)$ $ {\rm Re\,} \vec{\chi\,}(\infty)$, ${\rm Im} \vec{\tilde{\chi}\,}(t,0)$ $ = $ $\sin((\Omega -\Delta_r) t) $ $ {\rm Re} \vec{\chi\,}(\infty)$, 
 ${\rm Re\,} \vec{\tilde{\zeta}\,}_{0,1}(t,0)$ $ =$ $ \cos(\Omega t)$ $ {\rm Re\,} \vec{\zeta\,}_{0,1}(\infty)$ , ${\rm Im\,} \vec{\tilde{\zeta}\,}_{0,1}(t,0)$ $ = $ $ {\rm Re} \vec{\zeta\,}_{0,1}(\infty)\sin(\Omega  t)$, 
 ${\rm Re\,} \tilde{h}_{\pm}(t,0)$ $=$ $ \cos(\Delta_r t)$ $h_{\pm}(\infty)$ and ${\rm Im\,} \tilde{h}_{\pm}(t,0)$ $=$ $ \pm \sin(\Delta_r t)$ ${\rm Re\,}h_{\pm}(\infty)$. The flow of the imaginary parts ${\rm Im} \vec{\tilde{\chi}\,}$ 
 decouples from that of the real parts and of $h_0$ and of $H_1$. Thus it needs not to be considered.

%

\end{document}